\documentclass[ALICE,manyauthors]{cernphprep}
\usepackage[comma,square,numbers,sort&compress]{natbib}
\usepackage{hyperref}
\usepackage{lineno}
\usepackage{color}
\usepackage{caption}
\usepackage{multirow}

\usepackage{graphicx}
\usepackage{epstopdf}
\usepackage{amssymb}
\usepackage{xspace}
\usepackage{amsmath}
\usepackage{upgreek}
\usepackage{hyperref}
\usepackage[T1]{fontenc}

\usepackage{doi}

\newcommand{\snn}{$\sqrt{s_{\mathrm{NN}}}~$}
\newcommand{\s}{$\sqrt{s}~$}
\newcommand{\gvc}{GeV$/c$}
\newcommand{\kst}{$\mathrm{K^{*0}}$}

\newcommand{\kstba}{$\overline{\mathrm{K}}^{*0}$}
\newcommand{\kstf} {K$^{*}(892)^{0}~$}
\newcommand{\kstff} {K$^{*}(892)^{0}$}
\newcommand{\ph} {$\mathrm{\phi}~$}
\newcommand{\pha} {$\mathrm{\phi}$}
\newcommand{\phf} {$\mathrm{\phi(1020)}~$}
\newcommand{\phff} {$\mathrm{\phi(1020)}$}
\newcommand{\pt}{$p_{\mathrm{T}}~$}

\newcommand{\dndch}{$\langle\mathrm{d}N_{\mathrm{ch}}/\mathrm{d}\eta\rangle^{1/3}$}
\newcommand{\ptt}{\ensuremath{p_{\mathrm{T}}}\xspace}

\newcommand{\pb}{\mbox{Pb--Pb}\xspace}
\newcommand{\ppb}{\mbox{p--Pb}\xspace}

\newcommand{\kskm}{$\mathrm{K^{*0}/K}$}
\newcommand{\phikm}{$\mathrm{\phi/K}$}

\makeatletter{}
\usepackage{float} 

\begin{document}%
\begin{titlepage}

\PHyear{2019}
\PHnumber{249}      
\PHdate{30 October}  
  

\title{Evidence of rescattering effect in Pb--Pb collisions at the LHC
  through production of \kstf and \phf mesons}   
\ShortTitle{\kstf and \phf production in Pb--Pb collisions at
  \snn $=$ 5.02 TeV}
\Collaboration{ALICE Collaboration\thanks{See Appendix~\ref{app:collab} for the list of collaboration members}}
\ShortAuthor{ALICE Collaboration} 
\begin{abstract}
Measurements of \kstf and \phf resonance production in Pb--Pb and pp
collisions at \snn $=$ 5.02 TeV with 
the ALICE detector at the Large Hadron Collider are reported. The resonances are measured at midrapidity 
($|y|$ $<$ 0.5) via their hadronic decay channels and the transverse momentum (\ptt) distributions are obtained 
for various collision centrality classes up to \mbox{\pt $=$ 20 GeV$/c$}. The \ptt-integrated yield ratio \kstff$/$K
in \pb collisions shows significant suppression relative to pp collisions and decreases towards more central collisions.
In contrast, the \phff$/$K ratio does not show any suppression. Furthermore, the measured \kstff$/$K 
ratio in central \pb collisions is significantly suppressed with respect to the expectations based on a thermal model 
calculation, while the \phff$/$K ratio agrees with the model prediction. These measurements are an experimental 
demonstration of rescattering of \kstf~decay products in the hadronic phase of the collisions. The \kstff$/$K  
yield ratios in \pb and pp collisions 
are used to estimate the time duration between chemical and kinetic 
freeze-out, which is found to be $\sim$ 4--7 fm$/c$ for central collisions. The \ptt-differential ratios of \kstff$/$K, 
\phff$/$K, \kstff$/\pi$, \phff$/\pi$, $p/$\kstf and $p/$\phf are also presented for \pb and pp collisions 
at \snn $=$ 5.02 TeV. These ratios show that the rescattering effect is predominantly a low-\pt phenomenon. 
\end{abstract}
\end{titlepage}
\setcounter{page}{2}

\section{Introduction}\label{sec:intro}
Several measurements in high-energy heavy-ion collisions at the Large Hadron Collider (LHC)~\cite{Aamodt:2010pa,Aamodt:2010jd,ALICE:2011ab}
and the Relativistic Heavy Ion Collider (RHIC)~\cite{Adams:2005dq,Adcox:2004mh,Arsene:2004fa,Back:2004je,Gyulassy:2004zy,Shuryak:2008eq} have shown that a strongly-coupled Quark-Gluon Plasma (QGP) is formed that subsequently hadronizes.
Resonances, short lived hadrons that decay via strong interactions, play an 
important role in characterizing the properties of hadronic matter formed in heavy-ion collisions~\cite{Aggarwal:2010mt,Adams:2004ep,Adams:2006yu,ALICE:2018ewo,Abelev:2014uua,Acharya:2018qnp,Abelev:2008zk}. 
Several resonances have 
been observed in pp and nuclear collisions~\cite{Aggarwal:2010mt,Adams:2004ep,Abelev:2008zk,Adams:2006yu,ALICE:2018ewo,Abelev:2014uua,Acharya:2018qnp,Adam:2016bpr,Abelev:2012hy,Abelev:2008yz}: $f_{2}(1270)$, $\rho(770)^{0}$, $\Delta(1232)^{++}$, 
$f_{0}(980)$, $\rm{K}^{*}(892)^{0,\pm}$, $\Sigma(1385)$, $\Lambda(1520)$ and $\phi(1020)$ with lifetimes 
of the order of 
1.1 fm$/c$, 1.3 fm$/c$, 1.6 fm$/c$, 2.6 fm$/c$, 4.16 fm$/c$, 5.5 fm$/c$, 12.6 fm$/c$ and 46.3
fm$/c$, respectively~\cite{Tanabashi:2018oca}. The wide range of their lifetimes 
allows them to be good probes of the dynamics of the system formed in ultra-relativistic heavy-ion
collisions~\cite{Brown:1991kk,Bleicher:2002dm,Torrieri:2001ue,Johnson:1999fv,Markert:2008jc,Ilner:2017tab,Shapoval:2017jej}.  

In the hadronic phase of the evolution of the system formed in heavy-ion collisions, there are two important temperatures 
and corresponding timescales: the chemical freeze-out, when the inelastic collisions among the  constituents are expected 
to cease, and the later kinetic freeze-out, when all (elastic) interactions stop~\cite{Rapp:2000gy,Song:1996ik,Rafelski:2001hp}. 
If resonances decay before kinetic freeze-out, then their decay products are subject to hadronic rescattering that alters their 
momentum distributions. 
This leads to inability to reconstruct the parent resonance using the invariant mass technique, resulting in a decrease 
in the measured yield relative to the primordial resonance yield, i.e. the yield at chemical freeze-out. 
The fraction of resonances that cannot be recovered depends on the lifetime of the hadronic phase (defined as the time 
between chemical and kinetic freeze-out), 
the hadronic interaction cross section of resonance decay products, the particle density in the medium and the resonance 
phase space distributions. For example, a pion from a \kstf meson decay could scatter with another pion in the medium as
$\pi^{-}\pi^{+} \rightarrow \rho^{0}  \rightarrow \pi^{-}\pi^{+}$. At the same time, after the chemical freeze-out, pseudoelastic 
interactions could regenerate resonances in the medium, leading to an enhancement of their yields. For example, interactions 
like $\pi \rm{K} \rightarrow$ \kstf $\rightarrow \pi \rm{K}$ and K$^{-}$K$^{+}$ $\rightarrow$ \phf $\rightarrow$ K$^{-}$K$^{+}$
could happen until kinetic freeze-out. Hence, resonances are probes of the rescattering and regeneration processes during the 
evolution of the fireball from chemical to kinetic freeze-out. Indeed, transport-based model calculations  show that both 
rescattering and regeneration processes affect the final resonance yields~\cite{Singha:2015fia,Knospe:2015nva}. 
Thermal statistical models, 
which have successfully explained a host of particle yields in heavy-ion collisions across a wide range of center-of-mass
energies~\cite{Andronic:2008gu,Andronic:2017pug,Cleymans:1999st,Chatterjee:2015fua}, are able to explain the measured 
resonance yields only after including rescattering effects~\cite{Broniowski:2003ax,Rapp:2003ar}. 

In this paper, the measurement of the production of \kstf and \phf vector mesons  at midrapidity in \pb and pp  collisions at 
\snn $=$ 5.02 TeV is presented. Although both vector mesons have  similar masses, their lifetime differs by a factor of larger 
than 10. This aspect is exploited to establish the dominance of rescattering in central \pb  collisions at the LHC. The kaon 
and pion daughters of the short-lived \kstf $\rightarrow$ K$\pi$ rescatter with other hadrons in the medium. The magnitude of
the  effect is mainly determined by the pion-pion interaction cross section~\cite{Protopopescu:1973sh}, which is 
measured to be significantly larger (factor~5) than the total kaon-pion interaction cross section~\cite{Matison:1974sm}.  
The latter determines the magnitude of the regeneration effect~\cite{Bleicher:1999xi}. Thus with rescattering dominating over regeneration, 
the observable \kstf yields should decrease compared to the  primordial yields, and therefore, a suppression of the \kstff$/$K
yield ratio is expected in heavy-ion collisions relative to pp collisions. Furthermore, this ratio is expected to decrease with 
increase in system size, which is determined by the collision centrality (maximum for central collisions).  
 In contrast, because of a larger lifetime compared to that of the hadronic phase, the \phf meson yields are not expected to be affected by 
rescattering~\cite{Abelev:2014uua,Knospe:2015nva}. The \phf mesons are also expected not to be affected by the regeneration due to significantly lower KK cross section 
compared to K$\pi$ and $\pi\pi$ cross sections~\cite{Protopopescu:1973sh,Matison:1974sm}.
Hence the independence of the \phff$/$K yield ratio of the system size  will act as a baseline for corresponding 
\kstff$/$K measurements, thereby supporting the presence of the rescattering  effect in heavy-ion collisions. The lower \kstff$/$K 
yield ratio in \pb collisions compared to pp at the same \snn can then be used to estimate the time span between chemical 
and kinetic freeze-out in heavy-ion collisions.   Furthermore, due to the scattering of the decay products, the low-\ptt  \kstf
are less likely to escape the hadronic medium before decaying, compared to high-\ptt  \kstff~\cite{Knospe:2015nva}. This could alter the \kstf~\ptt 
spectra in \pb collisions compared to pp, while no such effect is
expected for \ph mesons. 
Therefore, studying \ptt-differential ratios of \kstf~and \phf mesons
with respect to other non-strange ($\pi$) and strange (K) mesons, and
baryons (p) in \pb and pp collisions will help to establish the \pt
dependence of rescattering effects and disentangle them from other
physics processes like radial flow that modifies the shapes of the \pt distributions at low and intermediate transverse momenta. 
In addition, the measurements at \snn $=$ 5.02 TeV are compared 
to results from Pb--Pb collisions at \snn $=$ 2.76
TeV~\cite{Abelev:2014uua, Adam:2017zbf}. Since production of particles 
and antiparticles is equal at midrapidity at LHC energies, the average
of the yields of \kstf and $\overline{\mathrm{K}}^*(892)^0~$ 
is presented in this paper and is denoted by the symbol \kst~unless specified otherwise. The \phf is denoted by the symbol 
\ph.

The paper is organized as follows: In section~\ref{sec:expSetup}, the detectors used in the analysis are briefly described. 
In section~\ref{sec:analysis}, the dataset, the analysis techniques, the procedure for extraction of the yields of \kst~and 
\ph mesons and the study of the systematic uncertainties are presented. In section~\ref{sec:results}, the yields obtained 
by invariant mass reconstruction of \kst~and \ph mesons as a function of transverse momentum in \pb and pp 
collisions at \snn $=$ 5.02 TeV, the \ptt-integrated ratios of \kst~and \ph relative to charged kaons, and \ptt-differential 
ratios relative to charged $\pi$, K and protons are reported. Finally, in section~\ref{sec:conc} the findings are summarized.

\section{Experimental apparatus} \label{sec:expSetup}
The measurements of \kst~and \ph meson production in pp and Pb--Pb collisions have been performed using the data 
collected by the ALICE detector in the year 2015. The details of the ALICE detector can be found in 
Ref.~\cite{Alessandro:2006yt,Aamodt:2008zz,Abelev:2014ffa}. 
So we briefly focus on the following main detectors used for this analysis.
 The forward V0 detector, a scintillator detector with a timing resolution less than 1 ns, is used for 
centrality selection, triggering and beam-induced background rejection. The V0 consists of two sub-detectors, V0A and V0C, 
placed at asymmetric positions, one on each side of the  interaction point with full azimuthal acceptance and cover the
pseudorapidity ranges 2.8 $<$ $\eta$ $<$ 5.1 and -3.7 $<$ $\eta$ $<$ -1.7, respectively. 
The centrality classes in Pb--Pb collisions 
are determined from the sum of the measured signal amplitudes in V0A and V0C, as discussed 
in Ref.~\cite{Adam:2015ptt,Aamodt:2010cz}. The collision time information is provided by T0 which consist of two arrays 
of Cherenkov counters T0A and T0C, positioned on both sides of the interaction point~\cite{Adam:2016ilk}. The Zero Degree
Calorimeter (ZDC) consists of two tungsten-quartz neutron and two
brass-quartz proton calorimeter placed at a distance of 113 m on both side of the interaction point. It is used to reject the background events and to measure the spectator nucleons.  

In the central barrel, the Inner Tracking System (ITS) and the Time Projection Chamber (TPC) are used for charged-particle 
tracking and primary collision vertex reconstruction. The ITS consists of  three sub-detectors of two layers each,  covering a 
central pseudorapidity  range $|\eta|$ $<$ 0.9: Silicon Pixel Detector (SPD), Silicon Drift Detector (SDD) and Silicon Strip 
Detector (SSD). The TPC is the main charged particle tracking detector, and has full azimuthal coverage in the pseudorapidity 
range $|\eta|$ $<$ 0.9. Along with track reconstruction, it also provides a measurement of the momentum and excellent particle
identification (PID). The TPC provides the measured specific energy loss (d$E/$d$x$) to identify the particles, especially in low 
momentum range ($p$ $<$ 1 GeV/$\it{c}$) where the d$E/$d$x$ of particles are well separated. To extend the particle 
identification to higher \ptt, the Time of Flight (TOF) detector is used in addition to the TPC information. 
The TOF is based on the Multigap Resistive Plate Chamber (MRPC) technology and measures the arrival times of particles with 
a resolution of the order of 80 ps. It covers a pseudorapidity range $|\eta|$ $<$ 0.9 and provides excellent PID capabilities in the 
intermediate \pt range by exploiting the time-of-flight information.

\section{Data sample and analysis details} \label{sec:analysis}
The pp data were collected using a minimum bias (MB) trigger. The logic for MB trigger requires  at least one hit in V0A or 
V0C and one hit in the central barrel detector SPD in coincidence with  the LHC bunch crossing
~\cite{Cortese:2004aa,Abbas:2013taa}. In pp collisions, a criterion based on the offline reconstruction of multiple primary 
vertices in the  SPD~\cite{Abelev:2014ffa} is applied to reduce the pileup, which is caused by multiple interactions in 
the same bunch crossing. The rejected pileup events are less than 1\% of the total events. The Pb--Pb data were also collected 
using a MB trigger with a logic that requires a coincidence of signals in V0A and V0C. The MB-triggered 
events are analyzed if they have a reconstructed collision vertex whose position along the beam axis 
($V_{z}$, $z$ is the longitudinal direction) is within 10 cm from the nominal interaction point in both pp and \pb collisions. 
Background events are rejected using the timing information from the Zero Degree Calorimeters (ZDCs) and V0 detectors.

The \pb~analysis is performed in 8 centrality classes defined in Ref.~\cite{Adam:2015ptt}: 0--10\%, 10--20\%, 20--30\%, 30--40\%, 40--50\%, 
50--60\%, 60--70\% and 70--80\%. The 0--10\% class corresponds to the most central Pb--Pb collisions, with 
small impact parameter, while the 70--80\% class corresponds to peripheral Pb--Pb collisions, with large impact parameter.
The total number of events that are analyzed after passing the event selection 
criteria are $\sim$110 million for pp and $\sim$30 million for Pb--Pb collisions. 
 Charged tracks are selected for analysis based on track selection criteria that ensure good track quality, 
as done in previous work~\cite{Adam:2017zbf}.
In particular, a track in the TPC is requested to have a minimum of 70 crossed rows (horizontal segments 
along the transverse readout plane of the TPC) out of a maximum possible 159~\cite{Acharya:2019yoi}. 
A \ptt-dependent selection criterion on the distance of closest approach to the collision vertex in the 
transverse ($xy$) plane (DCA$_{xy}$) and along the longitudinal direction (DCA$_z$)  is used to 
reduce the contamination from secondary charged particles coming from weakly decaying hadrons. 
 In addition to these selection criteria, tracks are required to have \pt $>$ 0.15 GeV$/c$ in both pp and 
Pb--Pb collisions. Charged particles are accepted in the pseudorapidity range  $|\eta|$ $\textless$ 0.8, 
which ensures a uniform acceptance. 
  
The particle identification exploits both the TPC and the TOF.  For K$^{*0}$ and $\phi$ reconstruction in  
Pb--Pb collisions,   charged particles are identified as pion or kaon if the mean specific energy loss 
($\langle$d$E/$d$x\rangle$) measured by the TPC  falls within  two standard deviations ($2\sigma_{\mathrm{TPC}}$) 
from the expected d$E/$d$x$ values for $\pi$ or K over the entire momentum range. If the TOF information 
is available for the track, in addition to the TPC, a TOF-based selection criterion $3\sigma_{\mathrm{TOF}}$ is 
applied over the measured momentum range, where $\sigma_{\mathrm{TOF}}$ is the standard deviation from 
the expected time-of-flight for a given species. 
These requirements help in reducing the background under the signal peak over a large momentum range 
and provide a better separation between signal and background with respect to TPC PID only.  For K$^{*0}$ 
reconstruction in pp collisions, the same PID selection criteria are applied to identify pion and kaon candidates 
as are used in Pb--Pb collisions. For the $\phi$ reconstruction in pp collisions, the kaon candidates are identified 
using a $6\sigma_{\mathrm{TPC}}$, $4\sigma_{\mathrm{TPC}}$ and $2\sigma_{\mathrm{TPC}}$ selection on 
the measured d$E/$d$x$ distributions in the momentum ranges $p$ $<$ 0.3 GeV$/c$, 0.3 $<$ $p$ $<$ 0.4 GeV$/c$ 
and $p$ $>$ 0.4 GeV$/c$, respectively. On top of this, the TOF-based selection criterion of  
$3\sigma_{\mathrm{TOF}}$ is applied over the entire measured momentum range in pp collisions if the 
TOF information is available. 

\subsection{Yield extraction, corrections and normalization} 
\label{sec:signal} 
The K$^{*0}$  and  $\phi$ resonances are reconstructed by  calculating the invariant mass of their decay products 
through the hadronic decay channels K$^{*0}(\overline{\rm{K}}^{*0}) \rightarrow \rm{K^{+}\pi^{-}}(\rm{K^{-}\pi^{+}})$ 
(Branching Ratio, BR = 66.666 $\pm$ 0.006\%~\cite{Tanabashi:2018oca}) and $\phi \rightarrow \rm{K^{+}K^{-}}$ 
(BR = 49.2 $\pm$ 0.5\%~\cite{Tanabashi:2018oca}), respectively. Oppositely  charged K and $\pi$ (or K) from 
the same event are paired to reconstruct the invariant mass distributions of K$^{*0}$($\phi$).  The K$\pi$ and KK 
pairs are selected in the rapidity range $|y|$ $<$ 0.5 in both pp and Pb--Pb collisions. The invariant mass distribution 
exhibits a signal peak and a large combinatorial background resulting from  the uncorrelated K$\pi$ (KK) pairs. The 
combinatorial background is estimated using a mixed-event technique in both collision systems. The mixed-event 
background is constructed by combining kaons from one event with the oppositely charged $\pi$(K) from different events 
for K$^{*0}(\phi)$. The events which are mixed are required to have similar characteristics. In Pb--Pb, two events are 
mixed if they belong to the same centrality class and the difference between the collision vertex position is 
$|\Delta V_{z}|$ $<$ 1 cm. In pp collisions, two events are mixed with a condition of  $|\Delta V_{z}|$ $<$ 1 cm 
and a difference in charged-particle density at midrapidity ($|\Delta{y}| < 0.5$) of less than 5. To minimize the statistical fluctuations 
in the background distribution, each event is mixed with five other ones. The invariant mass distribution from the mixed-event 
is normalized to the same-event oppositely-charged pair distribution in the mass region 1.1--1.3 (resp. 1.04--1.06) GeV$/c^{2}$ 
for K$^{*0}$(resp. $\phi$), which is away from the mass peak 
(6$\Gamma$ for K$^{*0}$ and 7$\Gamma$ for $\phi$, $\Gamma$ is the width of the resonance). 
 After the combinatorial background subtraction, the signal peak 
is observed on top of a residual background. The latter is due to the correlated K$\pi$ or KK pairs that originate from 
jets and from the misidentification of particles. It is shown in Ref.~\cite{Adam:2017zbf} that the residual background has 
a smooth dependence on mass and the shape of the background is well described by a second order polynomial 
\cite{Abelev:2014uua,Adam:2017zbf}. The 
invariant mass distributions after mixed-event background subtraction are fitted with a Breit-Wigner (resp. Voigtian) function 
for the signal peak of K$^{*0}$(resp. $\phi$) plus a second order polynomial for the residual background~\cite{Adam:2017zbf}. 
The Voigtian function is a convolution of a Breit-Wigner distribution and a Gaussian, where the width $\sigma$ of the 
Gaussian accounts for the mass resolution. The latter is \ptt-dependent and varies between 1 and 2 MeV$/c^{2}$. The 
raw yields are measured as a function of \pt for K$^{*0}$ and $\phi$  in pp collisions and  in various centrality classes 
in Pb--Pb collisions. A detailed description of the yield extraction procedure is given in Ref.~\cite{Adam:2017zbf}.

The measured yields are affected by the detector acceptance and reconstruction efficiency ($A\times \epsilon_{\mathrm{rec}}$). 
This is estimated by means of dedicated Monte Carlo simulations using
the PYTHIA  (PYTHIA 6 Perugia 2011 tune and PYTHIA 8 Monash 2013 tune)~\cite{Skands:2010ak,Skands:2014pea} and 
HIJING~\cite{Wang:1991hta}  event generators for pp and Pb--Pb collisions, respectively.  The generated particles are 
then propagated through the detector material using GEANT3~\cite{Brun:1082634}. The $A\times\epsilon_{\mathrm{rec}}$ 
is calculated as a function of \pt and is defined as the ratio of the reconstructed K$^{*0}$($\phi$) to the generated K$^{*0}$($\phi$), 
both within $|y| <$ 0.5. For the reconstruction of resonances, the same track and PID selection criteria are applied to the 
simulations as used in the analysis of the measured data. 
The $A \times\epsilon_{\mathrm{rec}}$  is calculated for K$^{*0}$($\phi$) that decay through the hadronic 
channel K$^{\pm}\pi^{\mp}$ (K$^{+}$K$^{-}$),  hence it does not include the correction for BR. 
In Pb--Pb collisions, the $A \times\epsilon_{\mathrm{rec}}$ has a weak centrality dependence and the raw yields are corrected using the 
$A \times\epsilon_{\mathrm{rec}}$ of the respective centrality class.

The procedure to correct the raw yields is given by 
\begin{equation}
  \frac{1}{N_{\mathrm{event}}}\frac{\mathrm{d}^{2}N}{\mathrm{d}y\mathrm{d}
    p_{\mathrm{T}}} =\frac{1}{N_{\mathrm{event}}^{\mathrm{acc}}}  \frac{\mathrm{d}^{2}N^{\mathrm{raw}}}{\mathrm{d}y\mathrm{d}
         p_{\mathrm{T}}} \frac{\epsilon_{\mathrm{trig}} ~.~\epsilon_{\mathrm{vert}}~.~ \epsilon_{\mathrm{sig}}} {(A \times 
         \epsilon_{\mathrm{rec}})~.~ \rm{BR}}.
       \label{yieldCorr}
\end{equation}
The raw yields are normalized to the  number of accepted events ($N_{\mathrm{event}}^{\mathrm{acc}}$) and corrected 
for $A \times \epsilon_{\mathrm{rec}} $, trigger efficiency ($\epsilon_{\mathrm{trig}}$), vertex reconstruction
efficiency ($\epsilon_{\mathrm{vert}}$), signal loss ($\epsilon_{\mathrm{sig}}$) and the BR of the 
decay channel. The yields in pp are normalized to the number of inelastic collisions with a trigger efficiency 
correction, $\epsilon_{\mathrm{trig}}$  = 0.757 $\pm$ 0.019~\cite{Loizides:2017ack}. The vertex reconstruction
efficiency  in pp collisions is found to be $\epsilon_{\mathrm{vert}}$ = 0.958. The signal loss correction factor 
$\epsilon_{\mathrm{sig}}$ is determined based on MC simulations as a function of \pt and  accounts for the resonance signal lost due to
trigger inefficiencies.  The $\epsilon_{\mathrm{sig}}$($p_{\mathrm{T}}$) correction is only significant for 
\pt $<$ 2.5 GeV$/c$ and has a value of less than 5\% both for  K$^{*0}$ and $\phi$ in pp collisions. In Pb--Pb
collisions, the yields of   K$^{*0}$ and $\phi$ in a given centrality class are normalized by the number of events
in the respective V0M (sum of V0A and V0C amplitude) event centrality class. The correction factors $\epsilon_{\mathrm{trig}}$, 
$\epsilon_{\mathrm{vert}}$ and $\epsilon_{\mathrm{sig}}$($p_{\mathrm{T}}$) are compatible with unity
 in the reported centrality classes in Pb--Pb collisions and hence are not used. 

\subsection{Systematic uncertainties} \label{sec:syst}
The systematic uncertainties in the measurement of K$^{*0}$ and $\phi$ yields in pp and Pb--Pb collisions 
are summarized in Tab.~\ref{tab_systematic}. The sources of systematic uncertainties are related to the 
yield extraction method, PID and track selection criteria, global tracking efficiency, the knowledge of the 
ALICE material budget and of the interaction cross section of hadrons in the detector material. The 
uncertainties are reported for three transverse momentum values, low, mid and high \ptt. For Pb--Pb 
collisions all the systematic uncertainties except the one related to the yield extraction are common in the
various centrality classes and  the values given in the table are averaged over all centralities. The yield 
extraction method includes the uncertainties due to variations of the fitting range, the choice of
combinatorial background estimation technique, normalization range and residual background shape. The
uncertainties due to yield extraction are estimated to be 
7.9--11.8$\%$ for K$^{*0}$ (resp. 2.4--3.5$\%$ for the $\phi$) in pp and 7.3--10.1$\%$ (resp. 1.9--4.9$\%$) in Pb--Pb collisions.
The PID systematic uncertainties varies between 2.1--6.9$\%$ (0.3--6.5$\%$) for K$^{*0}$ ($\phi$) in pp and Pb--Pb collisions.
The contribution to the uncertainty
from the global tracking efficiency is calculated from the corresponding values for single charged 
particles~\cite{Acharya:2019yoi}
and results in a  2.0--8.2\% uncertainty by combining the two charged tracks used in the invariant mass 
reconstruction of K$^{*0}$ and $\phi$. The contribution from variation of the track selection criteria is 1.0--5.5\%.  The systematic 
uncertainties due to the hadronic interaction cross section are estimated to be less than 2.8\% and contribute only
at low \ptt ($<$ 2 GeV$/c$).
The uncertainties in the 
description of the material budget of ALICE detector subsystems in GEANT3 (see Ref.~\cite{Abelev:2014laa} for 
details) give a contribution lower than 5.7$\%$ on the yields of K$^{*0}$ and $\phi$ in pp and \mbox{Pb--Pb} collisions. The 
material budget uncertainty is significant only at \pt $<$ 2 GeV$/c$ and negligible at higher 
$p_{\mathrm{T}}$. The total \ptt-dependent systematic uncertainties on the K$^{*0}$($\phi$) yields are estimated
to be 9.1--13.0\% (5.4--9.5$\%$) in pp collisions and 10.9--12.3$\%$ (6.4--9.2$\%$) in \mbox{Pb--Pb} collisions. The
common systematic uncertainties for different particles (global tracking efficiency, material budget and hadronic interaction) 
are cancelled out in calculating particle yield ratios like K$^{*0}/$K and $\phi/$K.

\begin{table}
\begin{center}
      \caption{Systematic uncertainties in the measurement of K$^{*0}$ and $\phi$ yields in pp and Pb--Pb 
      collisions at \snn $=$ 5.02 TeV. These uncertainties are shown for three transverse momentum values,
      low, mid and high \ptt. For Pb--Pb collisions all the systematic uncertainties except yield extraction are 
      common in various centrality classes and  the values given in the table are averaged over all centrality 
      classes.}
            \label{tab_systematic}
\scalebox{0.8}{
 \begin{tabular}{cccccccccccccccc}
\hline \hline
  &\multicolumn{7}{c}{Pb--Pb}                                               &    &\multicolumn{7}{c}{pp} \\
\hline 
Systematic variation        &  \multicolumn{3}{c} \kst & &\multicolumn{3}{c}\ph  &    &  \multicolumn{3}{c}\kst & &\multicolumn{3}{c}\ph\\
 \hline 
  &     \multicolumn{3}{c}{\ptt (GeV$/c$)}&&  \multicolumn{3}{c}{\ptt (GeV$/c$)}& &\multicolumn{3}{c}{\ptt (GeV$/c$)}&&\multicolumn{3}{c}{\ptt (GeV$/c$)} \\
      \cline{2-4} \cline{6-8} \cline{10-12} \cline{14-16}
	           &0.6&4.5&18 &&0.5&4.25&18& & 0.1&4.25&18& &0.5&4.25&18 \\
            \hline 
Yield extraction (\%)             &  7.3 &   7.5           &  10.1 & & 4.4 & 1.9 & 4.9  &    &  11.8   &7.9& 8.2  &           & 2.4 &3.5&3.5\\
Track selection (\%)               &  2.7  &  1.4              & 3.0  &&3.0 & 1.3  &  1.0  &    &   1.4  &1.0& 1.9         &       & 4.0   & 2.0 & 5.5    \\
Particle identification (\%)     &  5.4  &   3.0            & 5.0    &&1.0 &1.5 & 2.4   &    &   2.1   &3.2&  6.9        &       & 0.3   & 1.7& 6.5 \\
Global tracking efficiency (\%) &  4.7  &7.4                & 4.0 &&4.7 &8.2  & 3.1 &    &  2 .0   &3.1& 3.4           &   &2.0   &3.2&2.4    \\
Material budget (\%)           & 1.4 &0 & 0                  & &5.7 &0    &0        &     &  3.4   &0 & 0 && 5.7   &0&0\\
Hadronic Interaction (\%)      & 2.4 &0 & 0    &&1.3 &0    &0        &     &  2.8  &0& 0  && 1.3  &0&0\\
\hline
Total (\%)                                & 10.9  & 11.0& 12.3 &&9.2 &8.6  & 6.4  &&13.0 &9.1&  11.4 && 7.7   & 5.4 &9.5\\
\hline \hline
\end{tabular}
        }
      \end{center}

\end{table}

\section{Results and discussion}\label{sec:results}
\subsection{Transverse momentum spectra in pp and Pb--Pb collisions} \label{sec:ppcross}
\begin{figure}
      \begin{center}
        \includegraphics[scale=0.8]{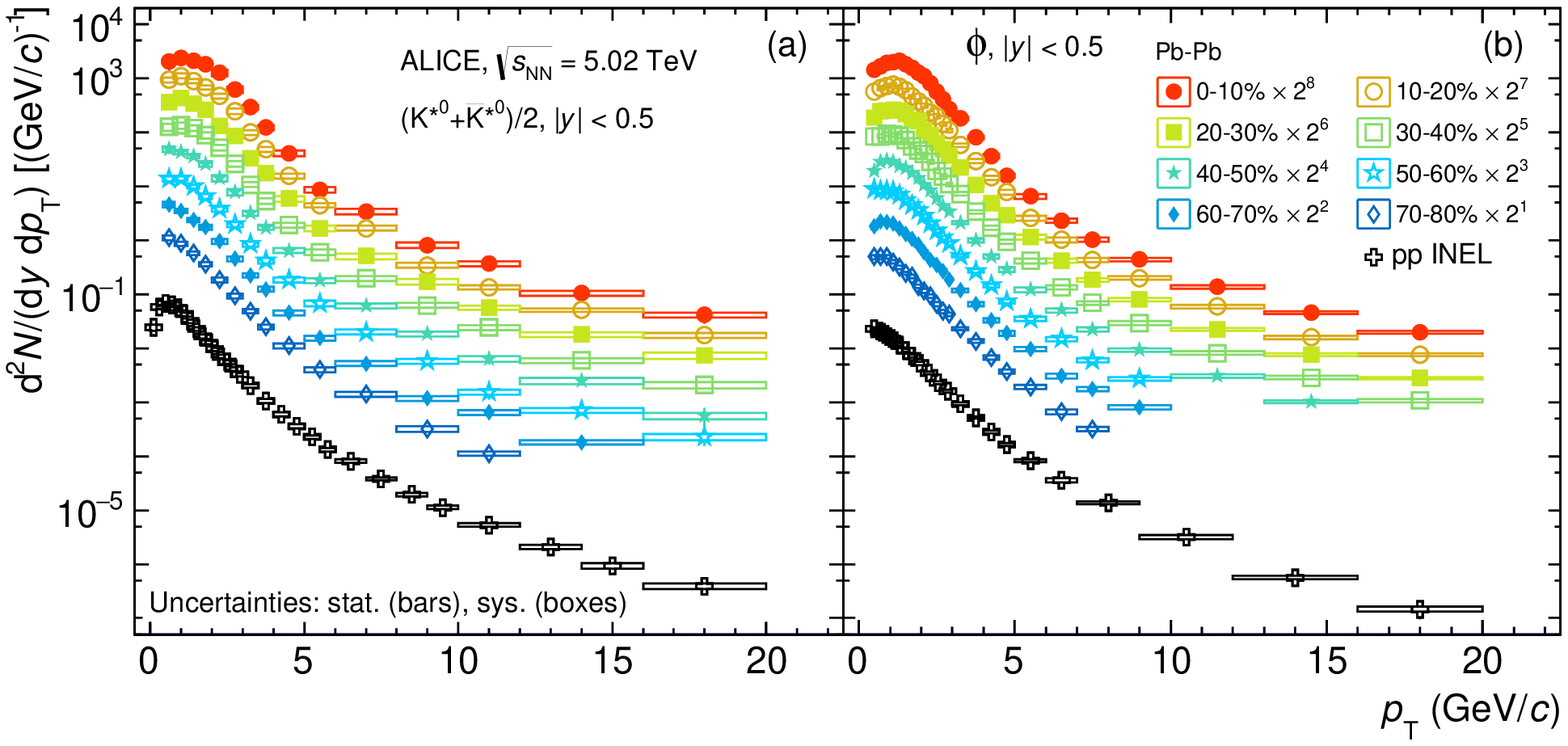}
       \caption{(Color online) 
The \ptt distributions of (a) K$^{*0}$ and (b) $\phi$ mesons in pp collisions and various centrality 
       	classes in Pb--Pb collisions at \snn $=$ 5.02 TeV. The values are plotted at the center of each bin. The statistical 
       	and systematic uncertainties are shown as bars and boxes, respectively.}
        \label{pbpb_spectrA}
      \end{center}
    \end{figure}

The \ptt distributions of the \kst~and \ph mesons for $|y|<0.5$, normalized to the number of events and 
corrected for efficiency, acceptance and branching ratio of the decay channel, are shown in Fig.~\ref{pbpb_spectrA}. 
The results for \pb collisions are presented for eight different centrality classes (0--10\% up to 70--80\% in 10\% wide 
centrality intervals) together with the results from inelastic pp collisions at the same energy. 

The \ptt-integrated particle yields have been extracted using the procedure described in 
Ref.~\cite{Abelev:2014uua,Adam:2017zbf}. The \ptt distributions are fitted with a L\'evy-Tsallis function
~\cite{Tsallis:1987eu,Abelev:2006cs} in pp and a Boltzmann-Gibbs blast-wave function
~\cite{Schnedermann:1993ws} in Pb--Pb collisions. The yields have been extracted from the data in the 
measured \ptt region and the fit functions have been used to extrapolate into the unmeasured (low and high \ptt) region. 
The low-\ptt extrapolation covers \ptt $<$ 0.4 GeV$/c$ for \kst  (\ph) and accounts for 8.6$\%$ (7.2$\%$) and
12.5\% (12.7\%)  of the total yield in the 0--10\% and 70--80\% centrality classes in \pb collisions, respectively. In pp collisions, the 
\kst~is measured in the range  0 $<$ \ptt $<$ 20 GeV$/c$. For the \ph meson, the low-\ptt extrapolation covers 
\ptt $<$ 0.4 GeV$/c$, accounting for 15.7\% of the total yield. The extrapolated fraction of the yield is negligible 
for \ptt $>$ 20 GeV$/c$. 

\subsection{Particle ratios}\label{sec:paricle}
\begin{figure}
      \begin{center}
        \includegraphics[scale=0.4]{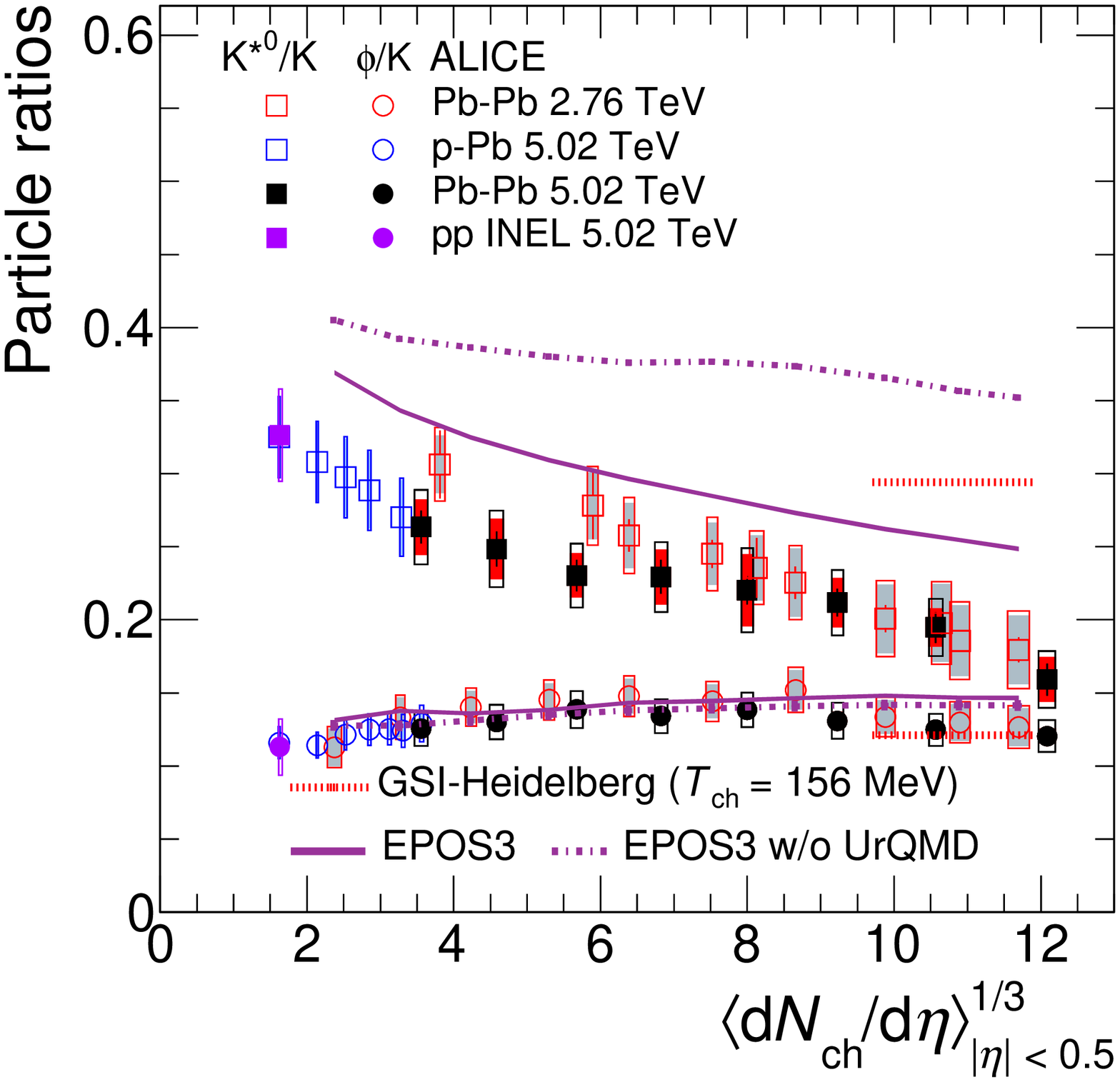}
        \caption{(Color online) \ptt-integrated particle yield ratios \kst$/$K$^{-}$ and $\phi/$K$^{-}$ as a function of 
        $\langle\mathrm{d}N_{\mathrm{ch}}/\mathrm{d}\eta\rangle^{1/3}$ measured at midrapidity in pp, p--Pb and Pb--Pb 
          collisions at \snn $=$ 5.02 TeV. For Pb--Pb collisions at \snn $=$ 2.76 TeV,  the \ph$/$K$^{-}$ values are taken from 
          Ref.~\cite{Abelev:2014uua} and the \kst$/$K$^{-}$ values are taken from Ref.~\cite{Adam:2017zbf}. The ratios for p--Pb
          collisions are taken from Ref.~\cite{Adam:2016bpr}. Statistical uncertainties (bars) are shown together with total 
          (hollow boxes) and charged-particle multiplicity-uncorrelated (shaded boxes)  systematic uncertainties. Thermal model calculations 
          with chemical freeze-out temperature $T_{\rm{ch}}$ $=$ 156 MeV for the most central Pb--Pb collisions
          ~\cite{Andronic:2017pug,Stachel:2013zma} are also shown. EPOS3 model predictions~\cite{Knospe:2015nva} of \kskm~and
	\phikm~ratios in Pb--Pb collisions are also shown as violet lines.}	
        \label{ratio_kstar_phi}
      \end{center}
    \end{figure}

Figure~\ref{ratio_kstar_phi} shows the \kskm~and \phikm~ratios as a function of \dndch~\cite{Aamodt:2010cz,Adam:2015ptt,Acharya:2019yoi}  for \pb collisions 
at \snn $=$ 2.76 ~\cite{Abelev:2014uua,Adam:2017zbf} and 5.02 TeV, \ppb collisions at \snn $=$ 5.02 TeV~\cite{Adam:2016bpr} and pp collisions 
at \mbox{\s $=$ 5.02 TeV}. The kaon yields in Pb--Pb at \snn $=$ 5.02 TeV are from Ref.~\cite{Acharya:2019yoi}. The \dndch~measured at midrapidity, is used here as a proxy for the system size. This is supported by the 
observation of the linear increase in the HBT radii with \dndch~\cite{Aamodt:2011mr,Lisa:2005dd}. The \kskm~ratio 
decreases for rising \dndch~while the \phikm~ratio is almost independent of  \dndch. The ratios exhibit a smooth trend 
across the different collision systems and collision energies studied. The \kskm~and \phikm~ratios in \pb collisions at 
\snn $=$ 2.76 and 5.02 TeV are in agreement within uncertainties.

The resonance yields are modified during the hadronic phase by rescattering (which would  reduce the measured 
yields) and regeneration (which would increase the measured yields).  The observed dependence of the \kskm~ratio 
on the charged-particle multiplicity is consistent with the behavior that would be expected if rescattering is the cause of the suppression. 
The fact that the \phikm~ratio does not exhibit suppression with charged-particle multiplicity suggests that the $\phi$, which has a lifetime 
an order of magnitude larger than that of the \kst, decays predominantly outside the hadronic medium. Theoretical 
estimates suggest that about 55\% of the of \kst~mesons with momentum $p=1$~\gvc, decay within 5~fm$/c$ of 
production (a typical estimate for the time between chemical and kinetic freeze-out in heavy-ion collisions
~\cite{Bleicher:2002dm,Bass:1999tu,Knospe:2015nva}), while only 7\% of \ph mesons with $p=1$~\gvc~ decay within 
that time. This supports the hypothesis that the experimentally observed decrease of the \kskm~ratio with charged-particle multiplicity 
is caused by rescattering.
A similar suppression has also been observed for $\rho^{0}/\pi$~\cite{Acharya:2018qnp} and 
$\Lambda^{*}/\Lambda$~\cite{ALICE:2018ewo} in central Pb--Pb collisions relative to peripheral Pb--Pb and pp collisions 
at \snn $=$ 2.76 TeV. In addition, the \kskm~ratio from thermal model calculations without rescattering effects and with 
chemical freeze-out temperature $T_{\rm{ch}}$ $=$ 156 MeV for the most central Pb--Pb collisions
~\cite{Andronic:2017pug,Stachel:2013zma} is found to be higher than the corresponding measurements, while the measured 
\phikm~ratio agrees with the thermal model predictions. The \kskm~and \phikm~ratios in Pb--Pb collisions are also 
compared to EPOS3 model calculations with and without a hadronic cascade phase modeled by UrQMD~\cite{Knospe:2015nva}. 
The EPOS3 model predictions shown in the figure are for Pb--Pb collisions at \snn $=$ 2.76 TeV but no significant qualitative 
differences are expected between the two energies. 
The EPOS3 generator with UrQMD reproduces the observed trend of the \kskm~and \phikm~ratios which further supports the experimental data.

    \begin{figure}
      \begin{center}
        \includegraphics[scale=0.45]{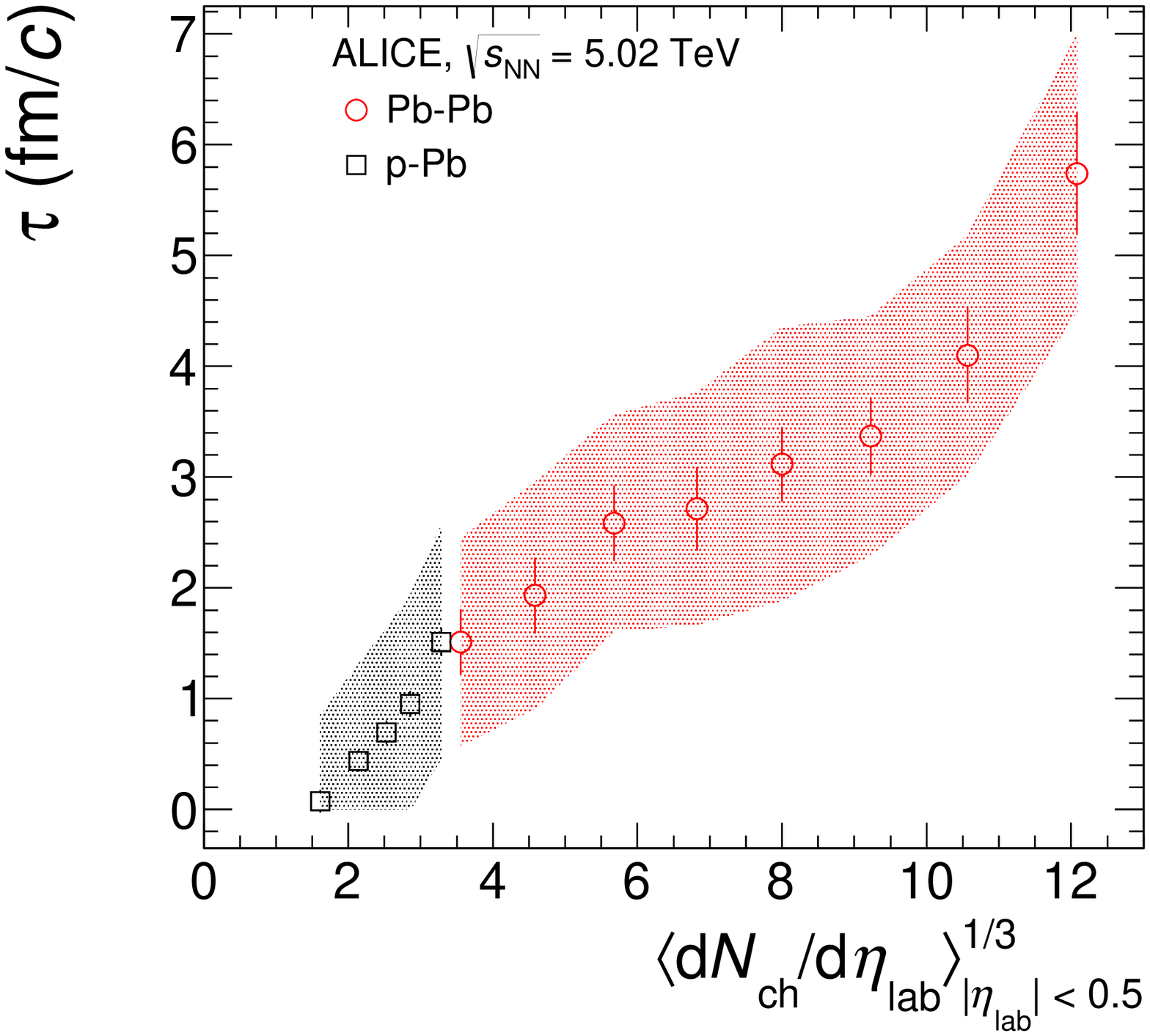}
        \caption{(Color online) Lower limit on the hadronic phase lifetime between chemical and kinetic freeze-out as a 
        function of  \dndch~in p--Pb~\cite{Adam:2016bpr} and Pb--Pb collisions at \snn $=$ 5.02 TeV. The bars and bands represent the
        statistical and systematic uncertainties, respectively,  propagated to the lifetime from the uncertainties associated 
        with the measured \kst$/$K ratios in \pb (\ppb) and pp collisions at \snn $=$ 5.02 TeV.}
        \label{hadron_life}
      \end{center}
    \end{figure}

The fact 
that \kst$/$K$^{-}$ decreases with increasing $\langle\mathrm{d}N_{\mathrm{ch}}/\mathrm{d}\eta\rangle^{1/3}$ 
implies that rescattering of the decay products of \kst~in the hadronic phase is dominant over $\mathrm{K}^{*0}$ 
regeneration. This suggests that $\mathrm{K}^{*0} \leftrightarrow \mathrm{K}\pi$ is not in balance. Hence in \pb the 
\kst$/$K$^{-}$  ratio can be used to get an estimate of the time between chemical and kinetic freeze-out, $\tau$, as, 
$[\mathrm{K}^{*0}/\mathrm{K}^{-}]_{kinetic}$ $=$ $[\mathrm{K}^{*0}/\mathrm{K}^{-}]_{chemical}$ 
$\times$ $e^{-\tau/\tau_{\rm{K}^{*0}}}$, where $\tau_{\rm{K}^{*0}}$ is the \kst~lifetime. Here, $\tau_{\rm{K}^{*0}}$~
is taken as 4.16 fm$/c$ ignoring any medium modification of the width of the invariant mass distribution of \kst. 
Furthermore, it is assumed that $[\mathrm{K}^{*0}/\mathrm{K}^{-}]_{chemical}$ is given by the values measured 
in pp collisions and the \pb collision data provides an estimate for $[\mathrm{K}^{*0}/\mathrm{K}^{-}]_{kinetic}$. 
This is equivalent to assuming that all \kst's that decay before kinetic freeze-out are lost due to rescattering effects 
and there is no regeneration effect between kinetic and chemical freeze-out which is supported by AMPT simulations
~\cite{Singha:2015fia}.  All the assumptions listed above lead to an estimate of  $\tau$ as a lower limit for the 
time span between chemical and kinetic freeze-outs. A decrease in the
\kst$/$K ratio with increasing multiplicity has previously also been
observed in in p--Pb collisions at \snn = 5.02 TeV
\cite{Adam:2016bpr}. This might indicate the presence of rescattering 
effect in high multiplicity p--Pb collisions and is suggestive of a finite lifetime of the hadronic phase. For comparison
we have also estimated the hadronic phase lifetime in p--Pb data. Figure~\ref{hadron_life} shows the results for $\tau$ 
boosted by a Lorentz factor ($\sim$ 1.65 for \ppb collisions and 1.75 for \pb collision) as a function of
$\langle\mathrm{d}N_{\mathrm{ch}}/\mathrm{d}\eta\rangle^{1/3}$. 
Neglecting higher order terms, the Lorentz factor is estimated as $\sqrt{1+(\langle p_{\mathrm{T}} \rangle/mc)^{2}}$.
Here $m$ is the rest 
mass of the resonance and $\langle p_{\mathrm{T}} \rangle$ is used as an approximation for $p$ for the 
measurements at midrapidity. The time interval between chemical and kinetic freeze-out increases with 
the system size as expected. For central \pb collisions at $\sqrt{s_{\mathrm{NN}}}$ $=$ 5.02 TeV, the lower 
limit of time between chemical and kinetic freeze-out is about 4--7 fm$/c$. This is of the same order of 
magnitude as the \kst~lifetime, 
but about an order of magnitude shorter than the \ph lifetime. A smooth increase of $\tau$ 
with system size from \ppb to \pb collisions is observed. The EPOS3
generator with UrQMD reproduces the increasing trend of $\tau$ with
multiplicity qualitatively~\cite{Knospe:2015nva}. If a constant chemical freeze-out temperature 
is assumed, then the increase of $\tau$ with multiplicity in \pb collisions corresponds to a decrease of the 
kinetic freeze-out temperature. This is in qualitative agreement with results from blast-wave fits to identified
particle \ptt distributions~\cite{Acharya:2019yoi}, which are interpreted as decrease in the kinetic freeze-out temperature 
from peripheral to central collisions.  

    \begin{figure}
      \begin{center}
        \includegraphics[scale=0.6]{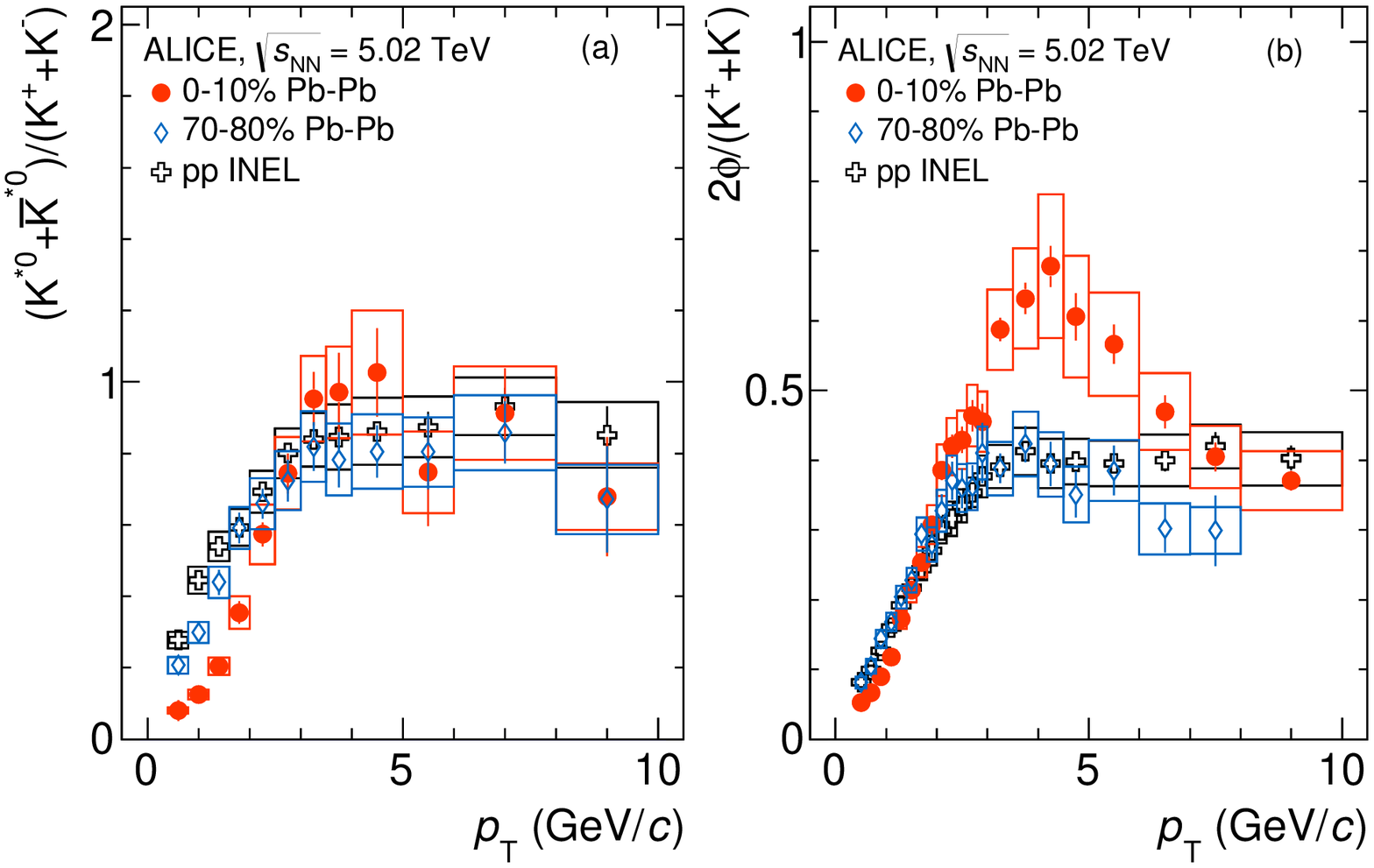}
        \caption{(Color online) Particle yield ratios (\kst~$+$
          \kstba)$/$(K$^{+}$ $+$ K$^{-}$) in panel (a) 
          and (2\pha)$/$(K$^{+}$ $+$ K$^{-}$) in panel (b), both as a function of \pt for
          centrality classes 0--10\% and 70--80\%  in \pb
          collisions at \snn $=$ 5.02 TeV.  For comparison, the
          corresponding ratios are also shown for inelastic pp
          collisions at \s $=$ 5.02 TeV. The statistical uncertainties
          are shown as bars and systematic uncertainties 
          are shown as boxes. In the text (\kst~$+$ \kstba),
          (K$^{+}$ $+$ K$^{-}$) are denoted by \kst~and K, respectively. }
       \label{ratio_K_phi_kstar_ptDiff}
      \end{center}
    \end{figure}

\begin{figure}[H]
     \begin{center}
       \includegraphics[scale=0.6]{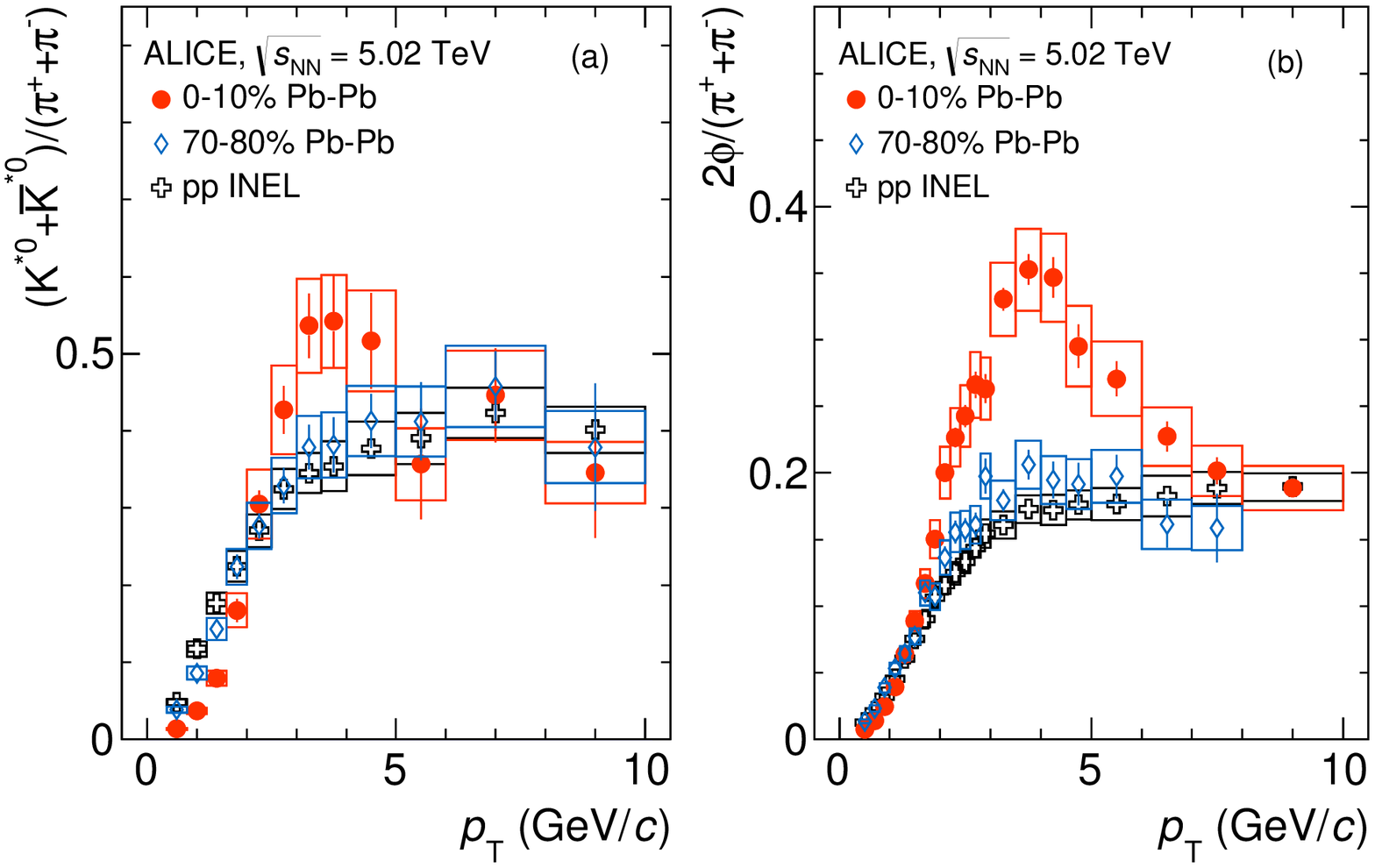}
       \caption{(Color online) Particle yield ratios (\kst~$+$
         \kstba)$/$($\pi^{+}$ $+$ $\pi^{-}$) in panel (a) 
          and (2\pha)$/$($\pi^{+}$ $+$ $\pi^{-}$) in panel (b), both as a function of \pt for 
          centrality classes 0--10\% and 70--80\%  in Pb--Pb 
          collisions at \snn $=$ 5.02 TeV.  For 
          comparison, the corresponding ratios are also shown for 
          inelastic pp collisions at \s $=$ 5.02 TeV. The statistical uncertainties 
          are shown as bars and systematic uncertainties 
          are shown as boxes. In the text (\kst~$+$ \kstba), 
          ($\pi^{+}$ $+$ $\pi^{-}$) are denoted by \kst~and $\pi$, respectively.}
       \label{ratio_pi_phi_kstar_ptDiff}
     \end{center}
   \end{figure}

\begin{figure}
     \begin{center}
       \includegraphics[scale=0.6]{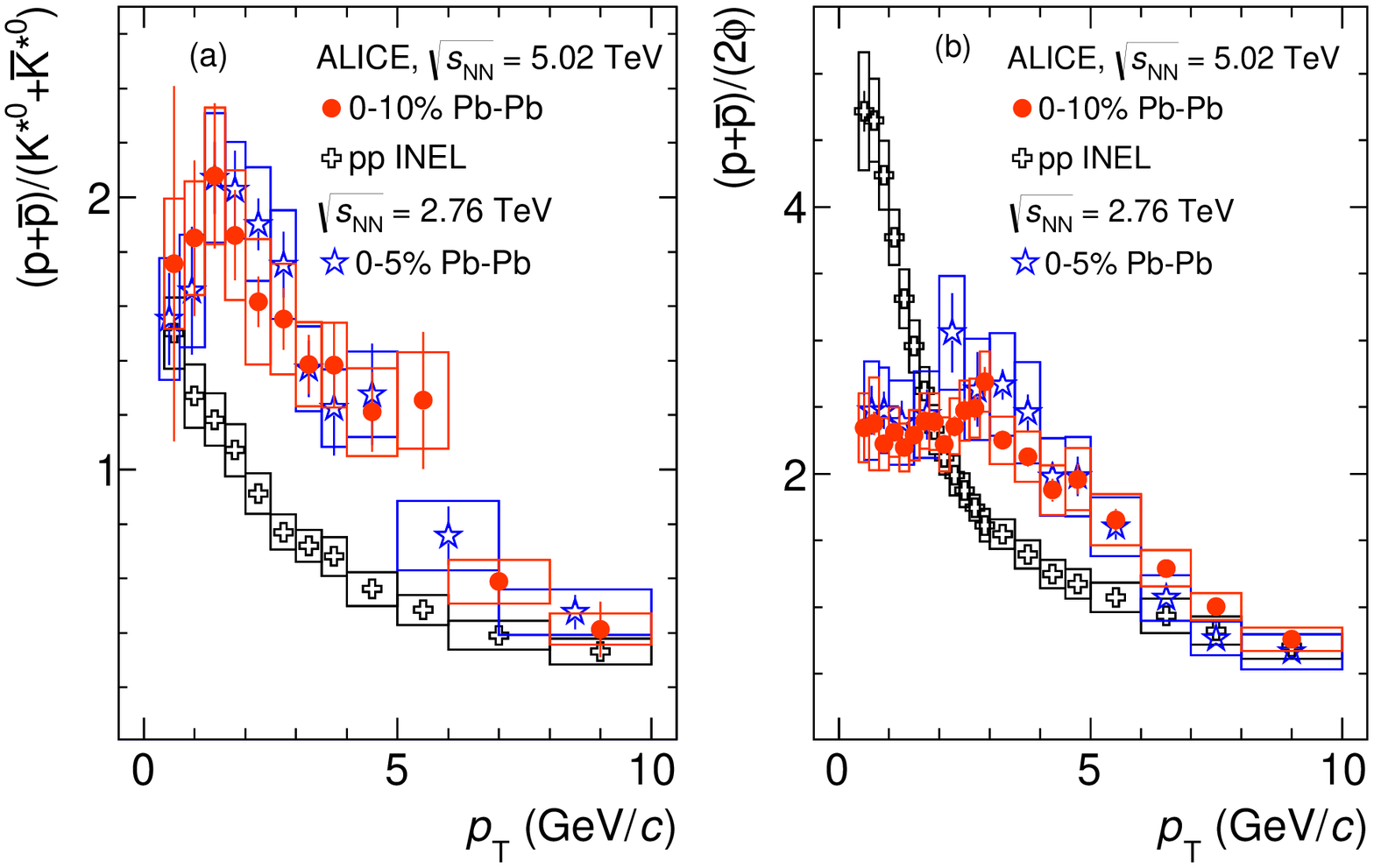}
       \caption{(Color online) Particle yield ratios  (p $+$ $\mathrm{\overline p}$)$/$(\kst~$+$ \kstba)  in panel (a) 
          and  (p $+$ $\mathrm{\bar p}$)$/$(2\pha) in panel (b), both as a function of \pt for 0--10\% central \pb 
          collisions and inelastic pp collisions at \mbox{\snn $=$ 5.02 TeV}.  For comparison, similar ratios are also 
          shown for 0--5\% central \pb collisions at \mbox{\snn $=$ 2.76 TeV~\cite{Adam:2017zbf}}. The statistical 
          uncertainties are shown as bars and systematic uncertainties
          are shown as boxes. In the text (\kst~$+$ \kstba) 
          and  (p $+$ $\mathrm{\overline p}$) are denoted by \kst~and p, respectively.}
        \label{ratio_p_kstar_phi_ptDiff}
    \end{center}
   \end{figure}

Further, to quantify the \ptt-dependence of the rescattering effect observed in \pb collisions, a set of 
$p_{\mathrm{T}}$-differential yield ratios was studied: K$^{*0}/$K, $\phi/$K, K$^{*0}/\pi$, $\phi/\pi$, 
$p/$K$^{*0}$ and $p/\phi$ as shown in Figs.~\ref{ratio_K_phi_kstar_ptDiff},~\ref{ratio_pi_phi_kstar_ptDiff}
and ~\ref{ratio_p_kstar_phi_ptDiff}. The choice of the ratios is
motivated by the following reasons: 
(a) the ratio of resonance yields relative to the ones of kaons and pions can shed light on the shapes of the \ptt distributions of mesons with different mass and quark content, and (b) the ratios of the proton yield with respect to the yields of the resonances allow comparisons among hadrons of similar mass, but different baryon number and quark content to be made.
For case (a), 
ratios in 0--10\%, 70--80\% Pb--Pb collisions and pp collisions at  \snn $=$ 5.02 TeV are compared. For case (b), 
ratios in 0--10\% Pb--Pb collisions and pp collisions at  \snn $=$ 5.02 TeV are compared with 0--5\% in \pb collisions 
at \snn $=$ 2.76 TeV. 
The ratios for 70--80\% in Pb--Pb collisions are closer to the corresponding results in pp collisions.
Noticeably, there 
are distinct differences between central and peripheral (pp) collisions in the ratios for \ptt below $\sim$ 2 \gvc~
and intermediate \ptt (between 2 and 6 \gvc)  but the ratios are consistent at higher \ptt~\cite{Adam:2017zbf}.  

At low \ptt, the K$^{*0}/$K
and K$^{*0}$/$\pi$  for central collisions are lower than in peripheral (pp) collisions, while the corresponding yield 
ratios for $\phi$ meson are comparable within the uncertainties. This observation is consistent with the suppression 
of \kst~yields due to rescattering in the hadronic phase. It demonstrates that rescattering affects low momentum 
particles. At intermediate \ptt, both ratios 
 show an enhancement for central \pb collisions 
relative to peripheral and pp collisions, which is more prominent for $\phi/$K, $\phi/\pi$ and K$^{*0}/\pi$. 
This is consistent with the presence of a larger radial flow in central collisions relative to peripheral and pp
collisions~\cite{Acharya:2019yoi}. Given that the masses of \kst~and \ph mesons are larger than
 those of the charged kaon and pion, the resonances experience a
 larger radial flow effect.  
In central \pb collisions, for \ptt below 5 \gvc, the $p/\phi$ ratio is observed to be independent of \ptt and the 
$p/$K$^{*0}$ ratio exhibits a weak \ptt-dependence within the uncertainties, in contrast to the decrease of both 
ratios with \ptt observed in pp collisions.
In turn, this suggests that the shapes of the \ptt distributions are similar for \kst, \ph and $p$ in this \ptt range. 
Although the quark contents are different, the masses of these hadrons are similar, indicating that this is the relevant quantity in determining spectra
shapes. This is consistent with expectations from hydrodynamic-based models~
\cite{Shen:2011eg, Minissale:2015zwa}. Within the uncertainties, the $p/$K$^{*0}$ and $p/\phi$ ratios for central \pb  
collisions at \snn = 5.02 TeV and 2.76 TeV~\cite{Adam:2017zbf} are constant at intermediate \ptt. This is consistent with the observation 
of similar order radial flow at both energies, obtained from the analysis of \ptt spectra of pions, kaons and 
protons~\cite{Acharya:2019yoi}.
For \ptt $>$ 6 \gvc,~ the
K$^{*0}/$K, $\phi/$K, K$^{*0}/\pi$, $\phi/\pi$, $p/$K$^{*0}$ and $p/\phi$ yield ratios in central collisions are 
similar to peripheral and pp collisions, indicating that fragmentation is the dominant hadron production mechanism 
in this \ptt region. This is consistent with previous measurements at \snn = 2.76 TeV~\cite{Adam:2017zbf}. 

   \section{Summary}\label{sec:conc}
   
The transverse momentum distributions of \kst~and \ph mesons have
been measured at midrapidity ($|y|<0.5$) 
for various collision centralities in \pb and inelastic pp collisions at \snn $=$ 5.02 TeV using the ALICE detector.  
The \kst~yields relative to charged kaons in \pb collisions show a suppression with respect to pp collisions, which 
increases with the system size, quantified using \dndch~measured at midrapidity. In contrast, no such suppression 
is observed for the \ph mesons. The lack of suppression for the \ph meson can be attributed to the fact that most of 
them decay outside the fireball because of its longer lifetime ($\tau_\phi$~= 46.3 $\pm$ 0.4 fm$/c$). Because of a 
shorter lifetime ($\tau_{\rm{K}^{*0}}$~ = 4.16 $\pm$ 0.05 fm$/c$), a significant number of produced \kst~decays in
the hadronic medium. The decay product(s) undergo interactions with other hadrons in the medium resulting in a significant 
change in their momentum, and no longer contributing to the \kst~signal reconstructed in the experiment.  Although 
both rescattering and regeneration are possible, the results presented here represent an experimental demonstration 
of the predominance of rescattering effects in the hadronic phase of the system produced in heavy-ion collisions. The effect of rescattering 
increases with the system size. Furthermore, the \kskm~yield ratios in central \pb collisions are significantly lower 
compared to the values from thermal model calculations without rescattering effects, while the measured \phikm~yield 
ratio agrees with the model calculation. This further corroborates the hypothesis that rescattering affects the
measured \kst~yields in \pb collisions. 
A lower limit for the lifetime of the hadronic phase is determined by using the \kst$/$K ratios 
in \pb and pp collisions at \snn $=$ 5.02 TeV. The lifetime, as expected, increases with system size. For central \pb
collisions, it is about 4--7 fm$/c$.  

The \ptt-differential yield ratios of  K$^{*0}/\pi$ and K$^{*0}/$K are studied in central \pb, peripheral \pb and pp 
collisions to understand the \ptt-dependence of the rescattering
effect. It is observed that rescattering dominantly affects the hadrons at \ptt $<$ 2 \gvc.  At intermediate \ptt (2--6 \gvc), the 
$\phi/$K, $\phi/\pi$, K$^{*0}/\pi$, $p/$K$^{*0}$ and $p/\phi$  yield  ratios are enhanced in central \pb collisions 
relative to peripheral \pb and pp collisions. In addition, the spectral shapes of \kst, \ph and $p$, which have comparable 
masses, are similar within the uncertainties for \ptt below 5 \gvc~in \pb collisions. These measurements demonstrate 
the effect of higher radial flow in central \pb collisions relative to peripheral \pb and pp collisions. 
A comparison of the 
$p/$K$^{*0}$ and $p/\phi$  ratios for central \pb collisions at \snn $=$ 5.02 and 2.76 TeV shows the constancy of the 
ratios with \ptt. This is consistent with the observation of comparable radial flow at \snn$=$  5.02 TeV and 2.76 TeV. 
For higher \ptt, above 6 \gvc, all the ratios agree within the uncertainties for central and peripheral \pb, and pp collisions, 
indicating that particle production via fragmentation at high transverse momenta is not significantly modified in the 
presence of a medium.   
\newenvironment{acknowledgement}{\relax}{\relax}
\begin{acknowledgement}
\section*{Acknowledgements}

The ALICE Collaboration would like to thank all its engineers and technicians for their invaluable contributions to the construction of the experiment and the CERN accelerator teams for the outstanding performance of the LHC complex.
The ALICE Collaboration gratefully acknowledges the resources and support provided by all Grid centres and the Worldwide LHC Computing Grid (WLCG) collaboration.
The ALICE Collaboration acknowledges the following funding agencies for their support in building and running the ALICE detector:
A. I. Alikhanyan National Science Laboratory (Yerevan Physics Institute) Foundation (ANSL), State Committee of Science and World Federation of Scientists (WFS), Armenia;
Austrian Academy of Sciences, Austrian Science Fund (FWF): [M 2467-N36] and Nationalstiftung f\"{u}r Forschung, Technologie und Entwicklung, Austria;
Ministry of Communications and High Technologies, National Nuclear Research Center, Azerbaijan;
Conselho Nacional de Desenvolvimento Cient\'{\i}fico e Tecnol\'{o}gico (CNPq), Financiadora de Estudos e Projetos (Finep), Funda\c{c}\~{a}o de Amparo \`{a} Pesquisa do Estado de S\~{a}o Paulo (FAPESP) and Universidade Federal do Rio Grande do Sul (UFRGS), Brazil;
Ministry of Education of China (MOEC) , Ministry of Science \& Technology of China (MSTC) and National Natural Science Foundation of China (NSFC), China;
Ministry of Science and Education and Croatian Science Foundation, Croatia;
Centro de Aplicaciones Tecnol\'{o}gicas y Desarrollo Nuclear (CEADEN), Cubaenerg\'{\i}a, Cuba;
Ministry of Education, Youth and Sports of the Czech Republic, Czech Republic;
The Danish Council for Independent Research | Natural Sciences, the VILLUM FONDEN and Danish National Research Foundation (DNRF), Denmark;
Helsinki Institute of Physics (HIP), Finland;
Commissariat \`{a} l'Energie Atomique (CEA), Institut National de Physique Nucl\'{e}aire et de Physique des Particules (IN2P3) and Centre National de la Recherche Scientifique (CNRS) and R\'{e}gion des  Pays de la Loire, France;
Bundesministerium f\"{u}r Bildung und Forschung (BMBF) and GSI Helmholtzzentrum f\"{u}r Schwerionenforschung GmbH, Germany;
General Secretariat for Research and Technology, Ministry of Education, Research and Religions, Greece;
National Research, Development and Innovation Office, Hungary;
Department of Atomic Energy Government of India (DAE), Department of Science and Technology, Government of India (DST), University Grants Commission, Government of India (UGC) and Council of Scientific and Industrial Research (CSIR), India;
Indonesian Institute of Science, Indonesia;
Centro Fermi - Museo Storico della Fisica e Centro Studi e Ricerche Enrico Fermi and Istituto Nazionale di Fisica Nucleare (INFN), Italy;
Institute for Innovative Science and Technology , Nagasaki Institute of Applied Science (IIST), Japanese Ministry of Education, Culture, Sports, Science and Technology (MEXT) and Japan Society for the Promotion of Science (JSPS) KAKENHI, Japan;
Consejo Nacional de Ciencia (CONACYT) y Tecnolog\'{i}a, through Fondo de Cooperaci\'{o}n Internacional en Ciencia y Tecnolog\'{i}a (FONCICYT) and Direcci\'{o}n General de Asuntos del Personal Academico (DGAPA), Mexico;
Nederlandse Organisatie voor Wetenschappelijk Onderzoek (NWO), Netherlands;
The Research Council of Norway, Norway;
Commission on Science and Technology for Sustainable Development in the South (COMSATS), Pakistan;
Pontificia Universidad Cat\'{o}lica del Per\'{u}, Peru;
Ministry of Science and Higher Education and National Science Centre, Poland;
Korea Institute of Science and Technology Information and National Research Foundation of Korea (NRF), Republic of Korea;
Ministry of Education and Scientific Research, Institute of Atomic Physics and Ministry of Research and Innovation and Institute of Atomic Physics, Romania;
Joint Institute for Nuclear Research (JINR), Ministry of Education and Science of the Russian Federation, National Research Centre Kurchatov Institute, Russian Science Foundation and Russian Foundation for Basic Research, Russia;
Ministry of Education, Science, Research and Sport of the Slovak Republic, Slovakia;
National Research Foundation of South Africa, South Africa;
Swedish Research Council (VR) and Knut \& Alice Wallenberg Foundation (KAW), Sweden;
European Organization for Nuclear Research, Switzerland;
Suranaree University of Technology (SUT), National Science and Technology Development Agency (NSDTA) and Office of the Higher Education Commission under NRU project of Thailand, Thailand;
Turkish Atomic Energy Agency (TAEK), Turkey;
National Academy of  Sciences of Ukraine, Ukraine;
Science and Technology Facilities Council (STFC), United Kingdom;
National Science Foundation of the United States of America (NSF) and United States Department of Energy, Office of Nuclear Physics (DOE NP), United States of America.   
      
\end{acknowledgement}
\bibliographystyle{utphys}
\bibliography{KStarPhiPbPb5p02TeV_letter}
    
\newpage
\appendix
    
\section{The ALICE Collaboration}
\label{app:collab}

\begingroup
\small
\begin{flushleft}
S.~Acharya\Irefn{org141}\And 
D.~Adamov\'{a}\Irefn{org94}\And 
A.~Adler\Irefn{org74}\And 
J.~Adolfsson\Irefn{org80}\And 
M.M.~Aggarwal\Irefn{org99}\And 
G.~Aglieri Rinella\Irefn{org33}\And 
M.~Agnello\Irefn{org30}\And 
N.~Agrawal\Irefn{org10}\textsuperscript{,}\Irefn{org53}\And 
Z.~Ahammed\Irefn{org141}\And 
S.~Ahmad\Irefn{org16}
% ...
\And
\renewcommand\labelenumi{\textsuperscript{\theenumi}~}

\section*{Affiliation notes}
\renewcommand\theenumi{\roman{enumi}}
\begin{Authlist}
\item \Adef{org*}Deceased
\item \Adef{orgI}Dipartimento DET del Politecnico di Torino, Turin, Italy
\item \Adef{orgII}M.V. Lomonosov Moscow State University, D.V. Skobeltsyn Institute of Nuclear, Physics, Moscow, Russia
\item \Adef{orgIII}Department of Applied Physics, Aligarh Muslim University, Aligarh, India
\item \Adef{orgIV}Institute of Theoretical Physics, University of Wroclaw, Poland
\end{Authlist}

\section*{Collaboration Institutes}
\renewcommand\theenumi{\arabic{enumi}~}
\begin{Authlist}
\item \Idef{org1}A.I. Alikhanyan National Science Laboratory (Yerevan Physics Institute) Foundation, Yerevan, Armenia
\item \Idef{org2}Bogolyubov Institute for Theoretical Physics, National Academy of Sciences of Ukraine, Kiev, Ukraine
\item \Idef{org3}Bose Institute, Department of Physics  and Centre for Astroparticle Physics and Space Science (CAPSS), Kolkata, India
\item \Idef{org4}Budker Institute for Nuclear Physics, Novosibirsk, Russia
\item \Idef{org5}California Polytechnic State University, San Luis Obispo, California, United States
\item \Idef{org6}Central China Normal University, Wuhan, China
\item \Idef{org7}Centre de Calcul de l'IN2P3, Villeurbanne, Lyon, France
\item \Idef{org8}Centro de Aplicaciones Tecnol\'{o}gicas y Desarrollo Nuclear (CEADEN), Havana, Cuba
\item \Idef{org9}Centro de Investigaci\'{o}n y de Estudios Avanzados (CINVESTAV), Mexico City and M\'{e}rida, Mexico
\item \Idef{org10}Centro Fermi - Museo Storico della Fisica e Centro Studi e Ricerche ``Enrico Fermi', Rome, Italy
\item \Idef{org11}Chicago State University, Chicago, Illinois, United States
\item \Idef{org12}China Institute of Atomic Energy, Beijing, China
\item \Idef{org13}Comenius University Bratislava, Faculty of Mathematics, Physics and Informatics, Bratislava, Slovakia
\item \Idef{org14}COMSATS University Islamabad, Islamabad, Pakistan
\item \Idef{org15}Creighton University, Omaha, Nebraska, United States
\item \Idef{org16}Department of Physics, Aligarh Muslim University, Aligarh, India
\item \Idef{org17}Department of Physics, Pusan National University, Pusan, Republic of Korea
\item \Idef{org18}Department of Physics, Sejong University, Seoul, Republic of Korea
\item \Idef{org19}Department of Physics, University of California, Berkeley, California, United States
\item \Idef{org20}Department of Physics, University of Oslo, Oslo, Norway
\item \Idef{org21}Department of Physics and Technology, University of Bergen, Bergen, Norway
\item \Idef{org22}Dipartimento di Fisica dell'Universit\`{a} 'La Sapienza' and Sezione INFN, Rome, Italy
\item \Idef{org23}Dipartimento di Fisica dell'Universit\`{a} and Sezione INFN, Cagliari, Italy
\item \Idef{org24}Dipartimento di Fisica dell'Universit\`{a} and Sezione INFN, Trieste, Italy
\item \Idef{org25}Dipartimento di Fisica dell'Universit\`{a} and Sezione INFN, Turin, Italy
\item \Idef{org26}Dipartimento di Fisica e Astronomia dell'Universit\`{a} and Sezione INFN, Bologna, Italy
\item \Idef{org27}Dipartimento di Fisica e Astronomia dell'Universit\`{a} and Sezione INFN, Catania, Italy
\item \Idef{org28}Dipartimento di Fisica e Astronomia dell'Universit\`{a} and Sezione INFN, Padova, Italy
\item \Idef{org29}Dipartimento di Fisica `E.R.~Caianiello' dell'Universit\`{a} and Gruppo Collegato INFN, Salerno, Italy
\item \Idef{org30}Dipartimento DISAT del Politecnico and Sezione INFN, Turin, Italy
\item \Idef{org31}Dipartimento di Scienze e Innovazione Tecnologica dell'Universit\`{a} del Piemonte Orientale and INFN Sezione di Torino, Alessandria, Italy
\item \Idef{org32}Dipartimento Interateneo di Fisica `M.~Merlin' and Sezione INFN, Bari, Italy
\item \Idef{org33}European Organization for Nuclear Research (CERN), Geneva, Switzerland
\item \Idef{org34}Faculty of Electrical Engineering, Mechanical Engineering and Naval Architecture, University of Split, Split, Croatia
\item \Idef{org35}Faculty of Engineering and Science, Western Norway University of Applied Sciences, Bergen, Norway
\item \Idef{org36}Faculty of Nuclear Sciences and Physical Engineering, Czech Technical University in Prague, Prague, Czech Republic
\item \Idef{org37}Faculty of Science, P.J.~\v{S}af\'{a}rik University, Ko\v{s}ice, Slovakia
\item \Idef{org38}Frankfurt Institute for Advanced Studies, Johann Wolfgang Goethe-Universit\"{a}t Frankfurt, Frankfurt, Germany
\item \Idef{org39}Fudan University, Shanghai, China
\item \Idef{org40}Gangneung-Wonju National University, Gangneung, Republic of Korea
\item \Idef{org41}Gauhati University, Department of Physics, Guwahati, India
\item \Idef{org42}Helmholtz-Institut f\"{u}r Strahlen- und Kernphysik, Rheinische Friedrich-Wilhelms-Universit\"{a}t Bonn, Bonn, Germany
\item \Idef{org43}Helsinki Institute of Physics (HIP), Helsinki, Finland
\item \Idef{org44}High Energy Physics Group,  Universidad Aut\'{o}noma de Puebla, Puebla, Mexico
\item \Idef{org45}Hiroshima University, Hiroshima, Japan
\item \Idef{org46}Hochschule Worms, Zentrum  f\"{u}r Technologietransfer und Telekommunikation (ZTT), Worms, Germany
\item \Idef{org47}Horia Hulubei National Institute of Physics and Nuclear Engineering, Bucharest, Romania
\item \Idef{org48}Indian Institute of Technology Bombay (IIT), Mumbai, India
\item \Idef{org49}Indian Institute of Technology Indore, Indore, India
\item \Idef{org50}Indonesian Institute of Sciences, Jakarta, Indonesia
\item \Idef{org51}INFN, Laboratori Nazionali di Frascati, Frascati, Italy
\item \Idef{org52}INFN, Sezione di Bari, Bari, Italy
\item \Idef{org53}INFN, Sezione di Bologna, Bologna, Italy
\item \Idef{org54}INFN, Sezione di Cagliari, Cagliari, Italy
\item \Idef{org55}INFN, Sezione di Catania, Catania, Italy
\item \Idef{org56}INFN, Sezione di Padova, Padova, Italy
\item \Idef{org57}INFN, Sezione di Roma, Rome, Italy
\item \Idef{org58}INFN, Sezione di Torino, Turin, Italy
\item \Idef{org59}INFN, Sezione di Trieste, Trieste, Italy
\item \Idef{org60}Inha University, Incheon, Republic of Korea
\item \Idef{org61}Institut de Physique Nucl\'{e}aire d'Orsay (IPNO), Institut National de Physique Nucl\'{e}aire et de Physique des Particules (IN2P3/CNRS), Universit\'{e} de Paris-Sud, Universit\'{e} Paris-Saclay, Orsay, France
\item \Idef{org62}Institute for Nuclear Research, Academy of Sciences, Moscow, Russia
\item \Idef{org63}Institute for Subatomic Physics, Utrecht University/Nikhef, Utrecht, Netherlands
\item \Idef{org64}Institute of Experimental Physics, Slovak Academy of Sciences, Ko\v{s}ice, Slovakia
\item \Idef{org65}Institute of Physics, Homi Bhabha National Institute, Bhubaneswar, India
\item \Idef{org66}Institute of Physics of the Czech Academy of Sciences, Prague, Czech Republic
\item \Idef{org67}Institute of Space Science (ISS), Bucharest, Romania
\item \Idef{org68}Institut f\"{u}r Kernphysik, Johann Wolfgang Goethe-Universit\"{a}t Frankfurt, Frankfurt, Germany
\item \Idef{org69}Instituto de Ciencias Nucleares, Universidad Nacional Aut\'{o}noma de M\'{e}xico, Mexico City, Mexico
\item \Idef{org70}Instituto de F\'{i}sica, Universidade Federal do Rio Grande do Sul (UFRGS), Porto Alegre, Brazil
\item \Idef{org71}Instituto de F\'{\i}sica, Universidad Nacional Aut\'{o}noma de M\'{e}xico, Mexico City, Mexico
\item \Idef{org72}iThemba LABS, National Research Foundation, Somerset West, South Africa
\item \Idef{org73}Jeonbuk National University, Jeonju, Republic of Korea
\item \Idef{org74}Johann-Wolfgang-Goethe Universit\"{a}t Frankfurt Institut f\"{u}r Informatik, Fachbereich Informatik und Mathematik, Frankfurt, Germany
\item \Idef{org75}Joint Institute for Nuclear Research (JINR), Dubna, Russia
\item \Idef{org76}Korea Institute of Science and Technology Information, Daejeon, Republic of Korea
\item \Idef{org77}KTO Karatay University, Konya, Turkey
\item \Idef{org78}Laboratoire de Physique Subatomique et de Cosmologie, Universit\'{e} Grenoble-Alpes, CNRS-IN2P3, Grenoble, France
\item \Idef{org79}Lawrence Berkeley National Laboratory, Berkeley, California, United States
\item \Idef{org80}Lund University Department of Physics, Division of Particle Physics, Lund, Sweden
\item \Idef{org81}Nagasaki Institute of Applied Science, Nagasaki, Japan
\item \Idef{org82}Nara Women{'}s University (NWU), Nara, Japan
\item \Idef{org83}National and Kapodistrian University of Athens, School of Science, Department of Physics , Athens, Greece
\item \Idef{org84}National Centre for Nuclear Research, Warsaw, Poland
\item \Idef{org85}National Institute of Science Education and Research, Homi Bhabha National Institute, Jatni, India
\item \Idef{org86}National Nuclear Research Center, Baku, Azerbaijan
\item \Idef{org87}National Research Centre Kurchatov Institute, Moscow, Russia
\item \Idef{org88}Niels Bohr Institute, University of Copenhagen, Copenhagen, Denmark
\item \Idef{org89}Nikhef, National institute for subatomic physics, Amsterdam, Netherlands
\item \Idef{org90}NRC Kurchatov Institute IHEP, Protvino, Russia
\item \Idef{org91}NRC «Kurchatov Institute»  - ITEP, Moscow, Russia
\item \Idef{org92}NRNU Moscow Engineering Physics Institute, Moscow, Russia
\item \Idef{org93}Nuclear Physics Group, STFC Daresbury Laboratory, Daresbury, United Kingdom
\item \Idef{org94}Nuclear Physics Institute of the Czech Academy of Sciences, \v{R}e\v{z} u Prahy, Czech Republic
\item \Idef{org95}Oak Ridge National Laboratory, Oak Ridge, Tennessee, United States
\item \Idef{org96}Ohio State University, Columbus, Ohio, United States
\item \Idef{org97}Petersburg Nuclear Physics Institute, Gatchina, Russia
\item \Idef{org98}Physics department, Faculty of science, University of Zagreb, Zagreb, Croatia
\item \Idef{org99}Physics Department, Panjab University, Chandigarh, India
\item \Idef{org100}Physics Department, University of Jammu, Jammu, India
\item \Idef{org101}Physics Department, University of Rajasthan, Jaipur, India
\item \Idef{org102}Physikalisches Institut, Eberhard-Karls-Universit\"{a}t T\"{u}bingen, T\"{u}bingen, Germany
\item \Idef{org103}Physikalisches Institut, Ruprecht-Karls-Universit\"{a}t Heidelberg, Heidelberg, Germany
\item \Idef{org104}Physik Department, Technische Universit\"{a}t M\"{u}nchen, Munich, Germany
\item \Idef{org105}Politecnico di Bari, Bari, Italy
\item \Idef{org106}Research Division and ExtreMe Matter Institute EMMI, GSI Helmholtzzentrum f\"ur Schwerionenforschung GmbH, Darmstadt, Germany
\item \Idef{org107}Rudjer Bo\v{s}kovi\'{c} Institute, Zagreb, Croatia
\item \Idef{org108}Russian Federal Nuclear Center (VNIIEF), Sarov, Russia
\item \Idef{org109}Saha Institute of Nuclear Physics, Homi Bhabha National Institute, Kolkata, India
\item \Idef{org110}School of Physics and Astronomy, University of Birmingham, Birmingham, United Kingdom
\item \Idef{org111}Secci\'{o}n F\'{\i}sica, Departamento de Ciencias, Pontificia Universidad Cat\'{o}lica del Per\'{u}, Lima, Peru
\item \Idef{org112}St. Petersburg State University, St. Petersburg, Russia
\item \Idef{org113}Stefan Meyer Institut f\"{u}r Subatomare Physik (SMI), Vienna, Austria
\item \Idef{org114}SUBATECH, IMT Atlantique, Universit\'{e} de Nantes, CNRS-IN2P3, Nantes, France
\item \Idef{org115}Suranaree University of Technology, Nakhon Ratchasima, Thailand
\item \Idef{org116}Technical University of Ko\v{s}ice, Ko\v{s}ice, Slovakia
\item \Idef{org117}Technische Universit\"{a}t M\"{u}nchen, Excellence Cluster 'Universe', Munich, Germany
\item \Idef{org118}The Henryk Niewodniczanski Institute of Nuclear Physics, Polish Academy of Sciences, Cracow, Poland
\item \Idef{org119}The University of Texas at Austin, Austin, Texas, United States
\item \Idef{org120}Universidad Aut\'{o}noma de Sinaloa, Culiac\'{a}n, Mexico
\item \Idef{org121}Universidade de S\~{a}o Paulo (USP), S\~{a}o Paulo, Brazil
\item \Idef{org122}Universidade Estadual de Campinas (UNICAMP), Campinas, Brazil
\item \Idef{org123}Universidade Federal do ABC, Santo Andre, Brazil
\item \Idef{org124}University of Cape Town, Cape Town, South Africa
\item \Idef{org125}University of Houston, Houston, Texas, United States
\item \Idef{org126}University of Jyv\"{a}skyl\"{a}, Jyv\"{a}skyl\"{a}, Finland
\item \Idef{org127}University of Liverpool, Liverpool, United Kingdom
\item \Idef{org128}University of Science and Techonology of China, Hefei, China
\item \Idef{org129}University of South-Eastern Norway, Tonsberg, Norway
\item \Idef{org130}University of Tennessee, Knoxville, Tennessee, United States
\item \Idef{org131}University of the Witwatersrand, Johannesburg, South Africa
\item \Idef{org132}University of Tokyo, Tokyo, Japan
\item \Idef{org133}University of Tsukuba, Tsukuba, Japan
\item \Idef{org134}Universit\'{e} Clermont Auvergne, CNRS/IN2P3, LPC, Clermont-Ferrand, France
\item \Idef{org135}Universit\'{e} de Lyon, Universit\'{e} Lyon 1, CNRS/IN2P3, IPN-Lyon, Villeurbanne, Lyon, France
\item \Idef{org136}Universit\'{e} de Strasbourg, CNRS, IPHC UMR 7178, F-67000 Strasbourg, France, Strasbourg, France
\item \Idef{org137}Universit\'{e} Paris-Saclay Centre d'Etudes de Saclay (CEA), IRFU, D\'{e}partment de Physique Nucl\'{e}aire (DPhN), Saclay, France
\item \Idef{org138}Universit\`{a} degli Studi di Foggia, Foggia, Italy
\item \Idef{org139}Universit\`{a} degli Studi di Pavia, Pavia, Italy
\item \Idef{org140}Universit\`{a} di Brescia, Brescia, Italy
\item \Idef{org141}Variable Energy Cyclotron Centre, Homi Bhabha National Institute, Kolkata, India
\item \Idef{org142}Warsaw University of Technology, Warsaw, Poland
\item \Idef{org143}Wayne State University, Detroit, Michigan, United States
\item \Idef{org144}Westf\"{a}lische Wilhelms-Universit\"{a}t M\"{u}nster, Institut f\"{u}r Kernphysik, M\"{u}nster, Germany
\item \Idef{org145}Wigner Research Centre for Physics, Budapest, Hungary
\item \Idef{org146}Yale University, New Haven, Connecticut, United States
\item \Idef{org147}Yonsei University, Seoul, Republic of Korea
\end{Authlist}
\endgroup
\end{document}